\documentclass{article}
\usepackage{amssymb}
\usepackage{amsmath}
\usepackage{amsfonts}
\usepackage{sw20ssur}
\usepackage{cite}
\usepackage{afterpage}

\newtheorem{theorem}{Theorem}
\newtheorem{acknowledgement}[theorem]{Acknowledgement}

\input{tcilatex}
\begin{document}

\title{Quantum Matching Pennies Game}
\author{Azhar Iqbal$^{\text{a,b}}$ and Derek Abbott$^{\text{a}}$ \\
%EndAName
$^{\text{a}}${\small School of Electrical \& Electronic Engineering, The
University of Adelaide, SA 5005, Australia.}\\
$^{\text{b}}${\small Centre for Advanced Mathematics and Physics,} {\small %
National University of Sciences \& Technology,}\\
{\small Campus of College of Electrical \& Mechanical Engineering, Peshawar
Road, Rawalpindi, Pakistan.}}
\maketitle
\tableofcontents

\begin{abstract}
A quantum version of the Matching Pennies (MP) game is proposed that is
played using an Einstein-Podolsky-Rosen-Bohm (EPR-Bohm) setting. We
construct the quantum game without using the state vectors, while
considering only the quantum mechanical joint probabilities relevant to the
EPR-Bohm setting. We embed the classical game within the quantum game such
that the classical MP game results when the quantum mechanical joint
probabilities become factorizable. We report new Nash equilibria in the
quantum MP game that emerge when the quantum mechanical joint probabilities
maximally violate the Clauser-Horne-Shimony-Holt form of Bell's inequality.
\end{abstract}

Keywords: quantum games, Nash equilibrium, EPR-Bohm experiment, quantum
probability

\section{Introduction}

A classical game \cite{Binmore,Rasmusen} can be considered an abstract
mathematical entity that is connected to the physical world \cite%
{AbbottShaliziDavies} in at least three recognizable ways: a) it describes a
strategic interaction among the participating players b) it is implemented
using a classical physical system that the players share to play the game c)
it is played in the presence of a referee who ensures that the participating
players abide by its rules.

Quantum games \cite{Mermin}-\cite{IqbalCheonAbbott} retain a) and c) but
they are distinguished from the classical games in that the physical system
used in the implementation of the game is quantum mechanical. This naturally
gives rise to the central question for the area of quantum games: How
quantum mechanical features of the shared physical system, used in the
physical implementation of the game, express themselves in terms of the
outcome/solution of the game?

For a faithful answer to this question it seems natural to establish, as the
first step, a correspondence between the classical features, or
classicality, of the shared physical system and the classical game and its
particular outcome. Establishing this correspondence paves the way for the
next step asking what impact it will have on the outcome/solution of the
game as the classical features of the shared physical system are replaced by
quantum features.

The physical system used in a two-party Einstein-Podolsky-Rosen-Bohm
(EPR-Bohm) experiment \cite%
{EPR,Bell,Bell(a),Bell1,Bell2,Aspect,Peres,Cereceda,WinsbergFine,Fine} is
known to have genuinely quantum features. This naturally motivates the use
of a two-party EPR-Bohm physical system to play a two-player quantum game.

Motivated by developing this approach towards quantum games, we proposed in
Ref. \cite{IqbalCheon} a scheme to play quantum games using EPR-Bohm
experiment. We reported that this scheme is able to construct genuine
quantum games from quantum mechanical probabilities only \cite{FootNote1}.
This is accomplished in the proposed scheme without referring to the quantum
mechanical state vectors, and with little reliance on the mathematical tools
of quantum mechanics.

We proposed this scheme for quantum games in view of Jarrett's position \cite%
{Jarrett} stating that the experimentally observed violations of Bell
inequalities in EPR-Bohm experiments are due to violations of the
conjunction of two probabilistic constraints---locality and completeness.
Jarrett concluded \cite{Jarrett} that \textquotedblleft the predictions of
quantum mechanics, in good agreement with the experimental results, satisfy
locality, but violate completeness.\textquotedblright\ Winsberg and Fine 
\cite{WinsbergFine,Fine} prefer the wording\textbf{\ }\textit{factorizability%
}\textbf{\ }for Jarrett's\textbf{\ }\textit{completeness}. We adopted
Winsberg and Fine's terminology in Ref. \cite{IqbalCheon} as well as in this
present paper. That is, the quantum features of EPR-Bohm experiments emerge
for non-factorizable joint probabilities.

By constructing quantum games from unusual non-factorizable joint
probabilities this scheme provides a unifying perspective for both quantum
and classical games, and also presents a more easily accessible analysis of
quantum games for researchers working outside the domain of quantum physics.

This scheme was developed for quantum games \cite{IqbalCheon} and applied it
to analyze the games \cite{Rasmusen} of Prisoner's Dilemma (PD), Stag Hunt,
and Chicken. For the PD game our analysis showed that, contrary to the
widely held belief, no new solution that is different from the classical
solution emerges when a quantum version of this game is constructed using an
EPR-Bohm setting.

However, within the same setting, for three-player PD \cite%
{IqbalCheonM,IqbalCheonAbbott} a new solution indeed emerges that is also
found to be Pareto-optimal \cite{Rasmusen}. Moreover, we showed that for the
two-player quantum Chicken game, new solution(s) arise for two identified
sets of quantum mechanical joint probabilities that maximally violate the
Clauser-Horne-Shimony-Holt (CHSH) sum of correlations \cite{Peres}.

The classical game of PD has a unique Nash equilibrium (NE) consisting of a
pair of identical pure strategies---and, in the two-player case, its quantum
version in the scheme using the EPR-Bohm setting, it does not generate a new
outcome. This motivates us, in the present paper, to study a quantum version
of a two-player game, within the same scheme, that has a unique mixed NE.
The well-known game of Matching Pennies (MP) \cite{Binmore,Rasmusen}
provides such an example.

Using the scheme based on EPR-Bohm experiments to play this game, we find
the impact on the solution of this game when the factorizability condition
on joint probabilities is dropped, while the conditions describing
normalization and locality are retained.

Another motive behind investigating the MP game, played using the EPR-Bohm
setting, is as follows. We notice that when multiple NE emerge in a
classical game, the analysis of its quantum version generates a separate set
of constraints on joint probabilities corresponding to that particular NE.
These constraints ensure that the classical game and its particular outcome
remains embedded within the quantum game. As the MP game has a unique mixed
classical NE, it presents an ideal situation to study how dropping the
factorizability condition on joint probabilities may change the outcome of
the game.

\section{Matching Pennies game}

In the game of MP each of the two players, henceforth labelled as Alice and
Bob, have a penny that each secretly flips to heads $\mathcal{H}$ or tails $%
\mathcal{T}$. No communication takes place between Bob and Alice and they
disclose their choices simultaneously to a referee, who organizes the game
and ensures that its rules are respected by the participating players.

If the referee finds that the pennies match (both heads or both tails), he
takes one dollar from Bob and gives it to Alice ($+1$ for Alice, $-1$ for
Bob). If the pennies do not match (one heads and one tails), the referee
takes one dollar from Alice and gives it to Bob ($-1$ for Alice, $+1$ for
Bob). As one player's gain is exactly equal to the other player's loss, the
game is zero-sum and is represented with the payoff matrix:

\begin{equation}
\begin{array}{c}
\text{Alice}%
\end{array}%
\begin{array}{c}
\mathcal{H} \\ 
\mathcal{T}%
\end{array}%
\overset{\overset{%
\begin{array}{c}
\text{Bob}%
\end{array}%
}{%
\begin{array}{cccc}
\mathcal{H} &  &  & \mathcal{T}%
\end{array}%
}}{\left( 
\begin{array}{cc}
(a_{1},b_{1}) & (a_{2},b_{2}) \\ 
(a_{3},b_{3}) & (a_{4},b_{4})%
\end{array}%
\right) },  \label{Matrix}
\end{equation}%
where we take $a_{1}=+1,$ $b_{1}=-1;$ $a_{2}=-1,$ $b_{2}=+1;$ $a_{3}=-1,$ $%
b_{3}=+1;$ and $a_{4}=+1,$ $b_{4}=-1$.

\subsection{Nash equilibrium}

It is well known that MP has no pure strategy Nash equilibrium \cite%
{Rasmusen} and instead has a unique mixed strategy NE. For completeness of
this paper we describe here how this is found. Consider repeated play of the
game in which $x$ and $y$ are the probabilities with which $\mathcal{H}$ is
played by Alice and Bob, respectively. The pure strategy $\mathcal{T}$ is
then played with probability $(1-x)$ by Alice, and with probability $(1-y)$
by Bob, and the players' payoff relations read

\begin{equation}
\Pi _{A,B}(x,y)=\left( 
\begin{array}{c}
x \\ 
1-x%
\end{array}%
\right) ^{T}\left( 
\begin{array}{cc}
(a_{1},b_{1}) & (a_{2},b_{2}) \\ 
(a_{3},b_{3}) & (a_{4},b_{4})%
\end{array}%
\right) \left( 
\begin{array}{c}
y \\ 
1-y%
\end{array}%
\right) .  \label{PayoffsClassicalGame}
\end{equation}%
A strategy pair $(x^{\star },y^{\star })$ is a NE when

\begin{equation}
\Pi _{A}(x^{\star },y^{\star })-\Pi _{A}(x,y^{\star })\geq 0,\text{ \ \ }\Pi
_{B}(x^{\star },y^{\star })-\Pi _{B}(x^{\star },y)\geq 0.  \label{NE}
\end{equation}%
For the matrix (\ref{Matrix}) these inequalities read $2(x^{\star
}-x)(2y^{\star }-1)\geq 0$ and $2(y^{\star }-y)(-2x^{\star }+1)\geq 0$ and
generate the strategy pair $(x^{\star },y^{\star })=(1/2,1/2)$ as the unique
NE of the game. At this NE the players' payoffs work out as

\begin{equation}
\Pi _{A}(1/2,1/2)=0=\Pi _{B}(1/2,1/2).  \label{payoffs at the NE}
\end{equation}

\subsection{Playing the game with 4 biased coins}

The first step in our quantization scheme for the MP game consists of
translating the game into a classical arrangement using a physical system
that involves $16$ joint probabilities. The arrangement we use consists of
two players sharing $4$ biased coins to play the game, assuming that the
referee has the means to set constraints on their biases.

The referee has $4$ coins and s/he marks them as $S_{1},S_{2},S_{1}^{\prime
},S_{2}^{\prime }$. S/he identifies $S_{1},S_{2}$ to be Alice's coins and $%
S_{1}^{\prime },S_{2}^{\prime }$ to be Bob's coins. In a run, the referee
hands over the $S_{1},S_{2}$ coins to Alice and the $S_{1}^{\prime
},S_{2}^{\prime }$ coins Bob. Alice's and Bob's strategies consist of
choosing one coin out of the two that each player receives in a run. The
pair of chosen coins in a run is one of the $(S_{1},S_{1}^{\prime }),$ $%
(S_{1},S_{2}^{\prime }),$ $(S_{2},S_{1}^{\prime }),$ $(S_{2},S_{2}^{\prime
}) $. The players return the two chosen coins to the referee who tosses them
together and records the outcome. The referee collects the $4$ coins ($2$
tossed and $2$ untossed) and repeats the same procedure over a large number
of runs.

Referee defines and makes public the players' payoff relations that depend
on a) the outcomes of a large number of tosses of $4$ biased coins, while $2$
coins are tossed in each run b) the players' strategies and c) the real
numbers defining the matrix of the game.

We now state that the statistical behavior of the $4$ biased coins,
expressed over a large number of tosses, is described by:

\begin{equation}
\begin{array}{c}
\text{Alice}%
\end{array}%
\begin{array}{c}
\underset{}{%
\begin{array}{c}
S_{1}%
\end{array}%
\begin{array}{c}
+1 \\ 
-1%
\end{array}%
} \\ 
\overset{}{%
\begin{array}{c}
S_{2}%
\end{array}%
\begin{array}{c}
+1 \\ 
-1%
\end{array}%
}%
\end{array}%
\overset{\overset{%
\begin{array}{c}
\text{Bob}%
\end{array}%
}{%
\begin{array}{cc}
\overset{%
\begin{array}{c}
S_{1}^{\prime }%
\end{array}%
}{%
\begin{array}{cc}
+1 & -1%
\end{array}%
} & \overset{%
\begin{array}{c}
S_{2}^{\prime }%
\end{array}%
}{%
\begin{array}{cc}
+1 & -1%
\end{array}%
}%
\end{array}%
}}{\left( 
\begin{tabular}{c|c}
$\underset{}{%
\begin{array}{cc}
p_{1} & p_{2} \\ 
p_{3} & p_{4}%
\end{array}%
}$ & $\underset{}{%
\begin{array}{cc}
p_{5} & p_{6} \\ 
p_{7} & p_{8}%
\end{array}%
}$ \\ \hline
$\overset{}{%
\begin{array}{cc}
p_{9} & p_{10} \\ 
p_{11} & p_{12}%
\end{array}%
}$ & $\overset{}{%
\begin{array}{cc}
p_{13} & p_{14} \\ 
p_{15} & p_{16}%
\end{array}%
}$%
\end{tabular}%
\right) },  \label{table}
\end{equation}%
where the $\mathcal{H}$ state of a coin is denoted by $+1$ and the $\mathcal{%
T}$ state by $-1$. The joint probabilities are factorizable for coins, that
is, one can find $4$ numbers $r,s,r^{\prime }$ and $s^{\prime }\in \lbrack
0,1]$ from which the joint probabilities can be obtained as

\begin{equation}
\begin{array}{llll}
p_{1}=rr^{\prime }, & p_{2}=r(1-r^{\prime }), & p_{3}=r^{\prime }(1-r), & 
p_{4}=(1-r)(1-r^{\prime }), \\ 
p_{5}=rs^{\prime }, & p_{6}=r(1-s^{\prime }), & p_{7}=s^{\prime }(1-r), & 
p_{8}=(1-r)(1-s^{\prime }), \\ 
p_{9}=sr^{\prime }, & p_{10}=s(1-r^{\prime }), & p_{11}=r^{\prime }(1-s), & 
p_{12}=(1-s)(1-r^{\prime }), \\ 
p_{13}=ss^{\prime }, & p_{14}=s(1-s^{\prime }), & p_{15}=s^{\prime }(1-s), & 
p_{16}=(1-s)(1-s^{\prime }),%
\end{array}
\label{factorizability}
\end{equation}%
where $r$ and $s$ are the probabilities of obtaining head for Alice's coins $%
S_{1}$ and $S_{2}$, respectively and, similarly, $r^{\prime }$ and $%
s^{\prime }$ are the probabilities of obtaining head for Bob's coins $%
S_{1}^{\prime }$ and $S_{2}^{\prime }$, respectively. In the following, we
call $r,s,r^{\prime },s^{\prime }$ the \emph{coin probabilities}.

\subsubsection{Payoff relations and Nash equilibrium\label{trans}}

The referee makes public and uses the following payoff relations:

\begin{equation}
\Pi _{A,B}(x,y)=\left( 
\begin{array}{c}
x \\ 
1-x%
\end{array}%
\right) ^{T}\left( 
\begin{array}{cc}
\Pi _{A,B}(S_{1},S_{1}^{\prime }) & \Pi _{A,B}(S_{1},S_{2}^{\prime }) \\ 
\Pi _{A,B}(S_{2},S_{1}^{\prime }) & \Pi _{A,B}(S_{2},S_{2}^{\prime })%
\end{array}%
\right) \left( 
\begin{array}{c}
y \\ 
1-y%
\end{array}%
\right) ,  \label{Q payoffs}
\end{equation}%
where $T$ is for transpose and $x$ and $y$ are the probabilities, definable
over a large number of runs, with which Alice and Bob choose $S_{1}$ and $%
S_{1}^{\prime }$, respectively. Also, that the referee defines

\begin{eqnarray}
\Pi _{A,B}(S_{1},S_{1}^{\prime }) &=&\tsum\nolimits_{i=1}^{4}(a,b)_{i}p_{i},%
\text{ \ \ \ \ \ }\Pi _{A,B}(S_{1},S_{2}^{\prime
})=\tsum\nolimits_{i=5}^{8}(a,b)_{i-4}p_{i},  \notag \\
\Pi _{A,B}(S_{2},S_{1}^{\prime })
&=&\tsum\nolimits_{i=9}^{12}(a,b)_{i-8}p_{i},\text{ \ \ }\Pi
_{A,B}(S_{2},S_{2}^{\prime })=\tsum\nolimits_{i=13}^{16}(a,b)_{i-12}p_{i}.
\label{QpayoffsParts}
\end{eqnarray}

It can be shown how, and under what circumstances, the payoff relations (\ref%
{Q payoffs}) produce the classical mixed-strategy game and result in the
classical NE. For the factorizable joint probabilities (\ref{factorizability}%
), obtained by a large number of coin tosses, the NE inequalities (\ref{NE})
read

\begin{eqnarray}
4(r-s)\left\{ y^{\star }(r^{\prime }-s^{\prime })+s^{\prime }-1/2\right\}
(x^{\star }-x) &\geq &0,  \notag \\
-4(r^{\prime }-s^{\prime })\left\{ x^{\star }(r-s)+s-1/2\right\} (y^{\star
}-y) &\geq &0.
\end{eqnarray}

At this stage, the referee sets the coin probabilities $r,s,r^{\prime
},s^{\prime }$ to be constrained as

\begin{equation}
r+s=1,\text{ \ \ }r^{\prime }+s^{\prime }=1,
\label{constraint on indepedent probs}
\end{equation}%
which, of course, then results in the strategy pair $(x^{\star },y^{\star
})=(1/2,1/2)$ to be the NE.

To obtain the players' payoffs at this NE, from Eqs.~(\ref{Q payoffs},\ref%
{constraint on indepedent probs}) we evaluate following quantities

\begin{eqnarray}
\Pi _{A}(S_{1},S_{1}^{\prime }) &=&(2r-1)(2r^{\prime }-1),\text{ \ \ }\Pi
_{A}(S_{1},S_{2}^{\prime })=(2r-1)(2s^{\prime }-1),  \notag \\
\Pi _{A}(S_{2},S_{1}^{\prime }) &=&(2s-1)(2r^{\prime }-1),\text{ \ \ }\Pi
_{A}(S_{2},S_{2}^{\prime })=(2s-1)(2s^{\prime }-1),
\end{eqnarray}%
from which the players' rewards at the NE of $(x^{\star },y^{\star
})=(1/2,1/2)$ are obtained as

\begin{equation}
\Pi _{A}(1/2,1/2)=0=\Pi _{B}(1/2,1/2).
\label{payoffs at NE when Prob are factor}
\end{equation}

We have thus translated the playing of MP game in an arrangement involving $%
16$ factorizable joint probabilities, obtained from a large number of tosses
performed on $4$ biased coins. We have found that, in order to guarantee
that factorizable joint probabilities result in the classical game, certain
constraints, given in (\ref{constraint on indepedent probs}), need to be
placed on the coin probabilities $r,s,r^{\prime },s^{\prime }$. This
translation paves the way for introducing the quantum mechanical joint
probabilities in the playing of this game, that may not be factorizable as
they are for classical coins.

\section{Quantum games using the EPR-Bohm setting}

We consider a quantum version of this game that is played using the EPR-Bohm
setting. This scheme for playing a quantum version of a two-player
two-strategy game was originally developed in Ref.\cite{IqbalCheon}. The
quantum game using the EPR-Bohm setting involves (refer to Fig.~1):

\begin{enumerate}
\item[a)] A large number of runs when, in a run, two halves of an EPR pair
originate from the same source and move in opposite directions.

\item[b)] One half is received by player Alice, while Bob receives the other
half. Alice and Bob are located at some distance from each other and are
unable to communicate between themselves.

\item[c)] The players, however, can communicate about their actions, which
they perform on their received halves, to the referee who organizes the game
and ensures that the rules of the game are followed.

\item[d)] The referee \cite{FootNote2} makes available two directions to
each player. Call Alice's two directions $S_{1}$ and $S_{2}$ and Bob's two
directions $S_{1}^{\prime }$ and $S_{2}^{\prime }$.

\item[e)] In a run, each player has to choose one of two directions at
his/her disposal and informs the referee of this choice.

\item[f)] After receiving information about the pair of directions, which
the players have chosen in a particular run, the referee rotates
Stern-Gerlach type detectors along the two chosen directions and performs a
quantum measurement.

\item[g)] The outcome of the quantum measurement \cite{FootNote3}, on
Alice's side, and on Bob's side of the Stern-Gerlach detectors, is either $%
+1 $ or $-1$.

\item[h)] Runs are repeated as the players receive a large number of halves
in pairs, when each pair comes from the same source.

\item[i)] The referee records the measurement outcomes for all runs, when in
each run each player chooses one of the two directions.

\item[j)] The referee defines a player's strategy, over a large number of
runs, to be a linear combination (with normalized and real coefficients) of
the two directions along which the measurement is performed.

\item[k)] The referee has payoff relations that s/he makes public at the
start of the game and announces rewards to the players after the completion
of runs.

\item[l)] The referee constructs these payoff relations in view of a) the
matrix (\ref{Matrix}) of the game being played, b) the list of players'
choices of directions over several runs, and c) the list of measurement
outcomes that the referee prepares using his/her Stern-Gerlach apparatus.
\end{enumerate}

\FRAME{fhFU}{2.9516in}{1.2246in}{0pt}{\Qcb{$S_{1}$ and $S_{2}$ are the two
directions that the referee assigns at the start of the game to Alice and,
in each run, Alice has to choose one direction. Over a large number of runs,
Alice chooses $S_{1}$ and $S_{2}$ with probabilities $x$ and $(1-x)$,
respectively. Similarly, the referee assigns two directions $S_{1}^{\prime }$
and $S_{2}^{\prime }$ to Bob at the start of the game and, in each run, he
has to choose one direction. Over a large number of runs, Bob chooses $%
S_{1}^{\prime }$ and $S_{2}^{\prime }$ with the probabilities $y$ and $(1-y)$%
, respectively.}}{\Qlb{Fig1}}{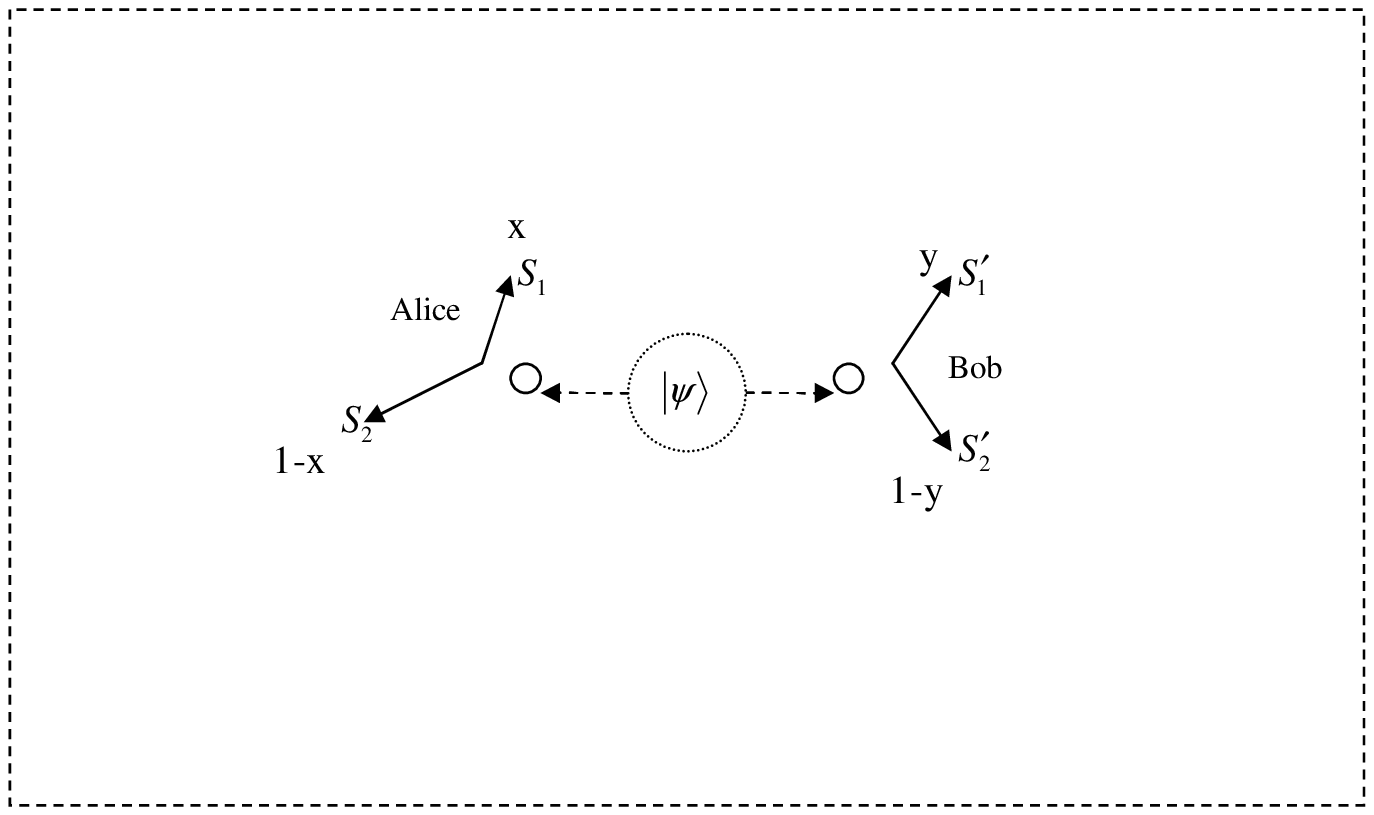}{\special{language "Scientific
Word";type "GRAPHIC";maintain-aspect-ratio TRUE;display "USEDEF";valid_file
"F";width 2.9516in;height 1.2246in;depth 0pt;original-width
5.4587in;original-height 3.2154in;cropleft "0.1989";croptop
"0.7367";cropright "0.7316";cropbottom "0.3661";filename
'draw1.eps';file-properties "XNPEU";}}

The translated MP game, using $4$ biased coins, allows one to express
players' payoff relations in terms of the $16$ joint probabilities. The
following Section shows that the physical system in the EPR-Bohm experiments
also involve $16$ joint probabilities, and thus the above translation
provides the natural route for playing a quantum MP game.

\subsection{Constraints on quantum mechanical joint probabilities}

The payoff relations (\ref{Q payoffs}) are defined in view of the fact that
the set of $16$ joint probabilities satisfy a number of constraints that are
imposed by the requirements of a) normalization, b) locality, and c)
factorizability.

In order to better appreciate the quantum mechanical probabilities, we
consider, for example, the situation when over all runs Alice chooses $S_{1}$
and Bob chooses $S_{2}^{\prime }$. Referee rotates Stern-Gerlach detectors
along these two directions and then, for example, referring to (\ref{table}) 
$p_{7}$ gives the probability of him/her obtaining $-1$ along Alice's $S_{1}$
direction and $+1$ along Bob's $S_{2}^{\prime }$ direction.

\subsubsection{Normalization}

Normalization says that when, for example, Alice chooses $S_{1}$ and Bob
chooses $S_{2}^{\prime }$ for all the runs, the only possible outcomes are $%
(+1,+1),$ $(+1,-1),$ $(-1,+1),$ $(-1,-1)$. The same is true for other pure
strategy pairs $(S_{1},S_{1}^{\prime }),$ $(S_{2},S_{1}^{\prime }),$ $%
(S_{2},S_{2}^{\prime })$:

\begin{equation}
\tsum\nolimits_{i=1}^{4}p_{i}=1=\tsum\nolimits_{i=5}^{8}p_{i},\text{ \ \ }%
\tsum\nolimits_{i=9}^{12}p_{i}=1=\tsum\nolimits_{i=13}^{16}p_{i}.
\label{normalization}
\end{equation}

\subsubsection{Locality}

The $16$ joint probabilities satisfy another set of constraints that are
obtained from the requirements stating that in a run:

a) Alice's outcome of $+1$ or $-1$ (obtained along $S_{1}$ or $S_{2}$) is
independent of whether Bob chooses $S_{1}^{\prime }$ or $S_{2}^{\prime }$ in
that run

b) Bob's outcome of $+1$ or $-1$ (obtained along $S_{1}^{\prime }$ or $%
S_{2}^{\prime }$) is independent of whether Alice chooses $S_{1}$ or $S_{2}$
in that run.

When translated in terms of joint probabilities, and referring to (\ref%
{table}), these requirements state that

\begin{eqnarray}
&&%
\begin{array}{cccc}
p_{1}+p_{2}=p_{5}+p_{6}, & p_{1}+p_{3}=p_{9}+p_{11}, & 
p_{9}+p_{10}=p_{13}+p_{14}, & p_{5}+p_{7}=p_{13}+p_{15}, \\ 
p_{3}+p_{4}=p_{7}+p_{8}, & p_{11}+p_{12}=p_{15}+p_{16}, & 
p_{2}+p_{4}=p_{10}+p_{12}, & p_{6}+p_{8}=p_{14}+p_{16}.%
\end{array}
\notag \\
&&  \label{locality constraints}
\end{eqnarray}

Quite often one finds in the literature the word `locality' to describe
these constraints. As can be seen, the possibility, described in (\ref%
{factorizability}), of writing $p_{i}$ for $1\leq i\leq 16$ in terms of $%
r,s,r^{\prime },s^{\prime }\in \lbrack 0,1]$ also assumes locality. Notice
that for a factorizable set of joint probabilities (\ref{factorizability})
the locality constraints (\ref{locality constraints}) always hold.

\subsubsection{Factorizability}

Eqs.~(\ref{factorizability}) state that the joint probabilities can be
written in terms of $r,s,r^{\prime },s^{\prime }\in \lbrack 0,1]$. If this
is the case then

\begin{equation}
r=p_{1}+p_{2},\text{ \ \ }s=p_{9}+p_{10},\text{ \ \ }r^{\prime }=p_{1}+p_{3},%
\text{ \ \ }s^{\prime }=p_{5}+p_{7},  \label{rsr's'}
\end{equation}%
and the Eqs.~(\ref{factorizability}) can be restated as

\begin{eqnarray}
p_{1} &=&(p_{1}+p_{2})(p_{1}+p_{3}),\text{ }%
p_{2}=(p_{1}+p_{2})(1-p_{1}-p_{3}),\text{ ...}  \notag \\
p_{16} &=&(1-p_{9}-p_{10})(1-p_{5}-p_{7})\text{.}  \label{factorizability2}
\end{eqnarray}%
The alert reader may notice that, in the writing of Eqs.~(\ref%
{factorizability},\ref{rsr's'}) and in the possibility of finding $%
r,s,r^{\prime },s^{\prime }\in \lbrack 0,1]$ that allows this, it is assumed
that joint probabilities satisfy the locality constraints (\ref{locality
constraints}).

\subsubsection{Cereceda's analysis}

We now refer to a result, reported by Cereceda \cite{Cereceda} stating that,
because of normalization (\ref{normalization}), half of the Eqs.~(\ref%
{locality constraints}) are redundant thus making eight among sixteen
probabilities $p_{i}$ independent. Cereceda has reported that a convenient
solution of the system (\ref{normalization}, \ref{locality constraints}), is
the one for which the set of variables:

\begin{equation}
\upsilon =\left\{
p_{2},p_{3},p_{6},p_{7},p_{10},p_{11},p_{13},p_{16}\right\} ,
\label{First set of probabilities}
\end{equation}%
is expressed in terms of the remaining set of variables:

\begin{equation}
\mu =\left\{ p_{1},p_{4},p_{5},p_{8},p_{9},p_{12},p_{14},p_{15}\right\} ,
\label{Second set of probabilities}
\end{equation}%
is given as

\begin{equation}
\begin{array}{l}
p_{2}=(1-p_{1}-p_{4}+p_{5}-p_{8}-p_{9}+p_{12}+p_{14}-p_{15})/2, \\ 
p_{3}=(1-p_{1}-p_{4}-p_{5}+p_{8}+p_{9}-p_{12}-p_{14}+p_{15})/2, \\ 
p_{6}=(1+p_{1}-p_{4}-p_{5}-p_{8}-p_{9}+p_{12}+p_{14}-p_{15})/2, \\ 
p_{7}=(1-p_{1}+p_{4}-p_{5}-p_{8}+p_{9}-p_{12}-p_{14}+p_{15})/2, \\ 
p_{10}=(1-p_{1}+p_{4}+p_{5}-p_{8}-p_{9}-p_{12}+p_{14}-p_{15})/2, \\ 
p_{11}=(1+p_{1}-p_{4}-p_{5}+p_{8}-p_{9}-p_{12}-p_{14}+p_{15})/2, \\ 
p_{13}=(1-p_{1}+p_{4}+p_{5}-p_{8}+p_{9}-p_{12}-p_{14}-p_{15})/2, \\ 
p_{16}=(1+p_{1}-p_{4}-p_{5}+p_{8}-p_{9}+p_{12}-p_{14}-p_{15})/2.%
\end{array}
\label{dependent probabilities}
\end{equation}

These relationships arise because the quantum mechanical joint probabilities
fulfill both the normalization condition (\ref{normalization}) as well as
the locality constraints (\ref{locality constraints}).

\subsubsection{CHSH inequality}

Notice that using (\ref{table}) the correlation $\left\langle
S_{1}S_{1}^{\prime }\right\rangle $, for example, can be found as

\begin{gather}
\left\langle S_{1}S_{1}^{\prime }\right\rangle =\Pr (S_{1}=1,S_{1}^{\prime
}=1)-\Pr (S_{1}=1,S_{1}^{\prime }=-1)  \notag \\
-\Pr (S_{1}=-1,S_{1}^{\prime }=+1)+\Pr (S_{1}=-1,S_{1}^{\prime }=-1)  \notag
\\
=p_{1}-p_{2}-p_{3}+p_{4}.
\end{gather}%
The correlations $\left\langle S_{1}S_{1}^{\prime }\right\rangle $, $%
\left\langle S_{1}S_{2}^{\prime }\right\rangle $, $\left\langle
S_{2}S_{1}^{\prime }\right\rangle $, and $\left\langle S_{2}S_{2}^{\prime
}\right\rangle $ can similarly be worked out. The CHSH sum of correlations
is then defined as

\begin{equation}
\Delta =\left\langle S_{1}S_{1}^{\prime }\right\rangle +\left\langle
S_{1}S_{2}^{\prime }\right\rangle +\left\langle S_{2}S_{1}^{\prime
}\right\rangle -\left\langle S_{2}S_{2}^{\prime }\right\rangle ,
\label{CHSH(a)}
\end{equation}%
and the CHSH inequality:

\begin{equation}
\left\vert \Delta \right\vert \leq 2,  \label{CHSH}
\end{equation}%
which holds for any theory of local hidden variables.

Cereceda has reported \cite{Cereceda} that there exist two sets of joint
probabilities that maximally violate the quantum prediction of the
Clauser-Holt-Shimony-Horne (CHSH) sum of correlations. The first set is
given as

\begin{equation}
\begin{array}{l}
p_{j}=(2+\sqrt{2})/8\text{ for all }p_{j}\in \mu , \\ 
p_{k}=(2-\sqrt{2})/8\text{ for all }p_{k}\in \upsilon ,%
\end{array}
\label{first set}
\end{equation}%
whereas the second set is given as

\begin{equation}
\begin{array}{l}
p_{j}=(2-\sqrt{2})/8\text{ for all }p_{j}\in \mu , \\ 
p_{k}=(2+\sqrt{2})/8\text{ for all }p_{k}\in \upsilon ,%
\end{array}
\label{second set}
\end{equation}%
where $\upsilon $ and $\mu $ are defined in (\ref{First set of probabilities}%
,\ref{Second set of probabilities}). That is, these two sets provide the
maximum absolute limit of $2\sqrt{2}$ for $\Delta _{QM}$.

\subsubsection{Constraints imposed by Cirel'son limit}

Now, alongside the constraints (\ref{constraint on joint probabilities})
there is another set of constraints on joint probabilities that are imposed
by the \emph{Cirel'son limit }\cite{Cirelson}, saying that the quantum
prediction of the CHSH sum of correlations $\Delta $, defined in (\ref%
{CHSH(a)}), is bounded in absolute value by $2\sqrt{2}$ i.e. $\left\vert
\Delta _{QM}\right\vert \leq 2\sqrt{2}$. Taking into account \cite{Cereceda}
the normalization condition (\ref{normalization}), the quantity $\Delta $ is
then equivalently expressed as

\begin{equation}
\Delta =2(p_{1}+p_{4}+p_{5}+p_{8}+p_{9}+p_{12}+p_{14}+p_{15}-2).
\label{delta}
\end{equation}%
In the following, the EPR setting, introduced in this Section, is used to
play the quantum version of the Matching Pennies game.

\section{Quantum Matching Pennies game}

Essentially, our quantum MP game corresponds when the $16$ joint
probabilities, that appear in the payoff relations (\ref{Q payoffs}), are
obtained using the EPR-Bohm setting, instead of using a large number of
tosses performed on biased coins.

The players' payoff relations in the quantum MP\ game, therefore, remain
exactly the same as they are defined and made public by the referee in Eq.~(%
\ref{Q payoffs}) for the translated game that uses factorizable joint
probabilities. Players' strategies also remain exactly the same as they are
in the classical game.

The referee is free to prepare any quantum pure or mixed bi-partite state
and to forward it to the players. S/he also fixes the $4$ available
directions at the start of the game (refer to Fig.~1) that cannot be changed
as the game progresses and large number of its runs are carried out. A
player's strategic choices do not go beyond choosing between the two
assigned directions.

\subsection{Embedding the classical game within the quantum game}

Referring to Eq.~(\ref{constraint on indepedent probs}) we recall that it
expresses the constraints on the coin probabilities. We also notice that the
factorizability, expressed by (\ref{factorizability}), permits one to write
the coin probabilities in terms of joint probabilities:

\begin{equation}
r=p_{1}+p_{2},\ \ s=p_{9}+p_{10},\ \ r^{\prime }=p_{1}+p_{3},\ \ s^{\prime
}=p_{5}+p_{7},
\end{equation}%
which allows us to rewrite the constraints (\ref{constraint on indepedent
probs}) on coin probabilities as

\begin{equation}
p_{1}+p_{3}+p_{5}+p_{7}=1,\text{ \ \ }p_{1}+p_{2}+p_{9}+p_{10}=1.
\label{constraint on joint probabilities}
\end{equation}%
This provides the the key for embedding the classical game within the
quantum game. S/he makes prior (experimental) arrangements in the EPR-Bohm
setup ensuring that the constraints (\ref{constraint on joint probabilities}%
) on joint probabilities hold during the whole course of playing the game 
\cite{FootNote4}. When this is the case the classical game remains embedded
within the corresponding quantum game in that the quantum game attains
classical interpretation with the joint probabilities becoming factorizable.

However, the joint probabilities that the EPR-Bohm setting can generate can
also be non-factorizable. This permits playing a quantum game in which the
constraints (\ref{constraint on joint probabilities}) hold, while the
factorizability condition on joint probabilities is dropped.

We now look at how dropping the factorizability condition for joint
probabilities affects the outcome of the game. With the constraints (\ref%
{constraint on joint probabilities}) continuing to hold, the referee can
then find a pair of NE strategies $(x^{\star },y^{\star })$ in the quantum
game using the inequalities (\ref{NE}) as usual. Because of non-factorizable
joint probabilities the strategy pair $(x^{\star },y^{\star })$ may be
different from the one which comes out for factorizable joint probabilities.

Notice that the rewards at the NE are identical to the ones given in (\ref%
{payoffs at the NE}). That is, when the $16$ joint probabilities become
factorizable, the NE and the players' payoffs become identical to the ones
obtained in the usual mixed strategy solution of the MP game. Also, the $16$
joint probabilities, even when they are non-factorizable and, therefore,
violate one or more of the set of Eqs.~(\ref{factorizability2}), will always
satisfy the normalization constraints (\ref{normalization}) as well as the
locality constraints (\ref{locality constraints}).

To be consistent with the standard setting \cite{Rasmusen} for playing a
two-player two-strategy game, the referee considers it reasonable to require
that in the EPR setting a player plays a pure strategy if s/he chooses the
same direction over all the runs and that s/he plays a mixed strategy if
s/he has a probability distribution with which s/he chooses between the two
directions at her/his disposal. However, identifying pure and mixed
strategies in such a way is not of much help as the payoff relations, which
referee uses to reward the players, generate the classical mixed strategy
game even when the players play `pure strategies.' This, however, remains
consistent with the known result in the area of quantum games stating that a
pure product initial state leads to the classical mixed strategy game.

\subsection{Nash equilibria in the quantum game}

We now find the NE that comes out from a set of non-factorizable (and thus
quantum mechanical) joint probabilities when the players' payoff relations
in the quantum game are obtained from the Eq.~(\ref{Q payoffs}). For the
inequalities defining the NE in the quantum game we obtain

\begin{gather}
\Pi _{A}(x^{\star },y^{\star })-\Pi _{A}(x,y^{\star })=[y^{\star }\left\{
\Pi _{A}(S_{1},S_{1}^{\prime })-\Pi _{A}(S_{2},S_{1}^{\prime })-\Pi
_{A}(S_{1},S_{2}^{\prime })+\Pi _{A}(S_{2},S_{2}^{\prime })\right\}  \notag
\\
+\left\{ \Pi _{A}(S_{1},S_{2}^{\prime })-\Pi _{A}(S_{2},S_{2}^{\prime
})\right\} ](x^{\star }-x)\geq 0,  \notag \\
\Pi _{B}(x^{\star },y^{\star })-\Pi _{B}(x^{\star },y)=[x^{\star }\left\{
\Pi _{B}(S_{1},S_{1}^{\prime })-\Pi _{B}(S_{1},S_{2}^{\prime })-\Pi
_{B}(S_{2},S_{1}^{\prime })+\Pi _{B}(S_{2},S_{2}^{\prime })\right\}  \notag
\\
+\left\{ \Pi _{B}(S_{2},S_{1}^{\prime })-\Pi _{B}(S_{2},S_{2}^{\prime
})\right\} ](y^{\star }-y)\geq 0,  \label{QNE}
\end{gather}%
where Eqs.~(\ref{QpayoffsParts}) and the matrix (\ref{Matrix}) gives

\begin{eqnarray}
\Pi _{A}(S_{1},S_{1}^{\prime }) &=&p_{1}-p_{2}-p_{3}+p_{4}=-\Pi
_{B}(S_{1},S_{1}^{\prime }),  \notag \\
\Pi _{A}(S_{1},S_{2}^{\prime }) &=&p_{5}-p_{6}-p_{7}+p_{8}=-\Pi
_{B}(S_{1},S_{2}^{\prime }),  \notag \\
\Pi _{A}(S_{2},S_{1}^{\prime }) &=&p_{9}-p_{10}-p_{11}+p_{12}=-\Pi
_{B}(S_{2},S_{1}^{\prime }),  \notag \\
\Pi _{A}(S_{2},S_{2}^{\prime }) &=&p_{13}-p_{14}-p_{15}+p_{16}=-\Pi
_{B}(S_{2},S_{2}^{\prime }),  \label{QPayoffsPartsExplicit}
\end{eqnarray}%
where the right sides of these equations express the fact that the quantum
game is a zero-sum game as is the classical game.

Using Eqs.~(\ref{dependent probabilities}) we eliminate the $8$
probabilities from the inequalities (\ref{QNE}) that gives the inequalities
for the NE in the quantum game in terms of the probabilities appearing in
the set (\ref{Second set of probabilities}):

\begin{gather}
\Pi _{A}(x^{\star },y^{\star })-\Pi _{A}(x,y^{\star })=2[y^{\star }\left\{
(1+p_{1}+p_{4})-(p_{5}+p_{8}+p_{9}+p_{12}+p_{14}+p_{15})\right\}  \notag \\
+(p_{5}+p_{8}+p_{14}+p_{15}-1)](x^{\star }-x)\geq 0,  \notag \\
\Pi _{B}(x^{\star },y^{\star })-\Pi _{B}(x^{\star },y)=-2[x^{\star }\left\{
(1+p_{1}+p_{4})-(p_{5}+p_{8}+p_{9}+p_{12}+p_{14}+p_{15})\right\}  \notag \\
+(p_{9}+p_{12}+p_{14}+p_{15}-1)](y^{\star }-y)\geq 0.  \label{QNE(a)}
\end{gather}

As some of the joint probabilities are constrained by (\ref{constraint on
joint probabilities}), using (\ref{dependent probabilities}) we rewrite
these constraints as

\begin{equation}
p_{9}+p_{15}=p_{12}+p_{14},\text{ \ \ }p_{5}+p_{14}=p_{8}+p_{15}.
\label{constraints(a)}
\end{equation}%
Now, adding the two equations in (\ref{constraints(a)}) and subtracting the
second from the first gives

\begin{equation}
p_{5}+p_{9}=p_{8}+p_{12},\text{ \ \ }%
p_{5}+p_{12}+2p_{14}=p_{8}+p_{9}+2p_{15},  \label{constraints(b)}
\end{equation}%
and we write

\begin{equation}
p_{12}=p_{5}+p_{9}-p_{8}\text{ and }p_{15}=p_{5}+p_{14}-p_{8},
\label{p12&p15}
\end{equation}%
in order to eliminate arbitrarily the probabilities $p_{12}$ and $p_{15}$
from the inequalities (\ref{QNE(a)}) to obtain

\begin{gather}
\Pi _{A}(x^{\star },y^{\star })-\Pi _{A}(x,y^{\star })=2[y^{\star }\left\{
(1+p_{1}+p_{4}+p_{8})-(3p_{5}+2p_{9}+2p_{14})\right\}  \notag \\
+\left\{ 2(p_{5}+p_{14})-1\right\} ](x^{\star }-x)\geq 0,  \notag \\
\Pi _{B}(x^{\star },y^{\star })-\Pi _{B}(x^{\star },y)=-2[x^{\star }\left\{
(1+p_{1}+p_{4}+p_{8})-(3p_{5}+2p_{9}+2p_{14})\right\}  \notag \\
+\left\{ 2(p_{5}-p_{8}+p_{9}+p_{14})-1\right\} ](y^{\star }-y)\geq 0.
\label{QNE(b)}
\end{gather}

The right sides of these inequalities involve six joint probabilities, which
we treat as `independent' and these are $%
p_{1},p_{4},p_{5},p_{8},p_{9},p_{14} $. These inequalities guarantee that
for a factorizable set of joint probabilities the classical mixed strategy
game of MP emerges.

\subsubsection{Nash equilibria for maximally entangled state}

Refer to the probability sets (\ref{first set},\ref{second set}) that
maximally violate the CHSH inequality. Probabilities in these sets are
non-factorizable as for both sets a solution for $r,$ $s,$ $r^{\prime },$ $%
s^{\prime }$ obtained from the Eqs.~(\ref{factorizability}) makes one or
more of the probabilities $r,$ $s,$ $r^{\prime },$ $s^{\prime }$ to be
negative or greater than one. This is also equivalent to stating that for
either of the sets (\ref{first set},\ref{second set}) one or more of the
equations (\ref{factorizability2}) does not hold, when $r,$ $s,$ $r^{\prime
},$ $s^{\prime }\in \lbrack 0,1]$ and the constraints (\ref{locality
constraints}) imposed by locality hold.

Now a natural question arising here is to ask if these two probability sets
can be used for the quantum game of MP. This will indeed be possible if for
each of these two sets the constraints given by (\ref{constraint on joint
probabilities}) hold ensuring that the classical MP game is embedded within
the quantum. For both the sets (\ref{first set},\ref{second set}) we find
that the constraint (\ref{constraint on joint probabilities}) hold, thus
these probability sets, maximally violating the CHSH sum of correlations,
can legitimately be used in the quantum MP game.

For the first set (\ref{first set}) the inequalities (\ref{QNE(b)}) work out
as

\begin{eqnarray}
\Pi _{A}(x^{\star },y^{\star })-\Pi _{A}(x,y^{\star }) &=&\sqrt{2}(-y^{\star
}+1)(x^{\star }-x)\geq 0,  \notag \\
\Pi _{B}(x^{\star },y^{\star })-\Pi _{B}(x^{\star },y) &=&-\sqrt{2}%
(-x^{\star }+1)(y^{\star }-y)\geq 0,  \label{QNE(c)}
\end{eqnarray}%
which give the strategy pairs $(1,0)\ $and $(1,1)$ as NE. At the strategy
pair $(1,0)$ the players' payoffs are obtained from Eqs. (\ref{Q payoffs},%
\ref{QPayoffsPartsExplicit}) as $\Pi _{A}(1,0)=1/\sqrt{2}=-\Pi _{B}(1,0)$
whereas at the strategy pair $(1,0)$ the players' payoffs are obtained to be
the same i.e. $\Pi _{A}(1,1)=1/\sqrt{2}=-\Pi _{B}(1,1)$.

Similarly, for the second set (\ref{second set}) the NE inequalities (\ref%
{QNE(b)}) are

\begin{eqnarray}
\Pi _{A}(x^{\star },y^{\star })-\Pi _{A}(x,y^{\star }) &=&\sqrt{2}(y^{\star
}-1)(x^{\star }-x)\geq 0,  \notag \\
\Pi _{B}(x^{\star },y^{\star })-\Pi _{B}(x^{\star },y) &=&-\sqrt{2}(x^{\star
}-1)(y^{\star }-y)\geq 0,  \label{QNE(d)}
\end{eqnarray}%
giving the strategy pairs $(0,1)\ $and $(1,1)$ as the NE. At the strategy
pair $(0,1)$ the players' payoffs work out as $\Pi _{A}(0,1)=-1/\sqrt{2}%
=-\Pi _{B}(0,1)$ whereas at the strategy pair $(1,1)$ the players' payoffs
are obtained as the same i.e. $\Pi _{A}(1,1)=-1/\sqrt{2}=-\Pi _{B}(1,1)$.

\section{Discussion}

This paper is motivated by the observation that by having a unique mixed NE
the classical MP game offers an opportunity for seeing more clearly how
dropping the factorizability condition on joint probabilities may affect
this unique NE, which emerges for factorizable joint probabilities in the
quantization scheme based on EPR-Bohm experiments.

Notice that in the scheme based on EPR-Bohm experiments the referee's role
is significantly increased as compared to other schemes for playing quantum
games. This is because s/he is free to provide any pair of directions to
each player and makes quantum measurement(s) on any pure or mixed bi-partite
states. The available options for the players are, therefore, reduced in
comparison to what is the case in other quantization schemes \cite%
{EWL,MarinattoWeber}, and they have exactly the same options as in the
classical game. In a classical two-player two-strategy game each player can
play a linear combination (with real and normalized coefficients) of two
pure strategies and this remains exactly the same in the our scheme for
playing a two-player quantum game.

Joint probabilities in EPR-Bohm experiments, performed on entangled
bipartite states, are known to become non-factorizable when players make
their strategic choices along certain pairs of directions. This provides the
opportunity to look at the possible new outcomes of the game that
non-factorizable joint probabilities may generate. In the quantization
scheme based on EPR-Bohm experiments the constraints placed on probabilities
guarantee that the classical game remains embedded within the quantum game,
while probabilities may become non-factorizable.

By constructing quantum games directly from quantum probabilities the
suggested approach contributes towards an understanding and potential use of
quantum probabilities in the area of game theory. That is, the question
addressed in this paper asks whether quantum probabilities have more to
offer to game theory. The answer to this we find is `yes'.

The possibility that CHSH inequality can be rephrased in terms of two-player
cooperative games has been reported in literature. In Ref. \cite{Cleve}
Cleve et al. have reported a game based on CHSH inequality in which the
maximum probability of winning the classical game is $3/4$ whereas, using a
quantum strategy, the players can win this game with probability $0.85$,
which, as they show using Cirel'son's limit, is optimal. Also, in Ref. \cite%
{Cheon} Cheon has reported a quantum game in which both players maximize a
quantity (utility) defined from spin projections of two particles which they
share, whereas the payoff operators are the measurement operator for
EPR-Bohm experiment. Cheon then finds the NE of the quantum game that
rewards the players far better for particles with maximum entanglement
compared to when the particles are uncorrelated.

Both of these studies show that EPR-Bohm experiments can be translated into
special games. The contribution of the quantization scheme developed in Ref. 
\cite{IqbalCheon}, and that of the present paper, however, is that it shows
that, along with the reported possibility of translating EPR-Bohm
experiments as special games, one can in fact quantize any two-player game
using the framework of EPR-Bohm experiments. Secondly, that we can analyze
our quantum game using the non-factorizable property of quantum mechanical
joint probabilities.

Nonfactorizability is known \cite{Jarrett,WinsbergFine,Fine} to be a
necessary but insufficient condition for the violation of Bell's inequality,
the CHSH form of which we consider here. That is, a set of $16$\ joint
probabilities that violates Bell's inequality will always be
non-factorizable, whereas one can find a set of joint probabilities that is
non-factorizable and still does not violate the CHSH form of Bell's
inequality. This known result has the following implications when it is
considered in our scheme for playing quantum games using EPR-Bohm
experiments: As a new solution of the game, which emerges because of
dropping the factorizability condition, the relevant joint probabilities may
not violate the Bell's inequality (in its CHSH form)---only those outcomes
of the quantum game are to be considered to have a \textit{bona fide}
quantum aspect \cite{CheonTsutsui} for which the corresponding set of joint
probabilities violates the CHSH form of Bell's inequality. The NE of the
quantum game for which the Bell's inequality is not violated will,
therefore, have a pseudoclassical aspect.

Using Bell's inequality one can identify the pseudoclassical domain from the
quantum domain as follows. With the constraints (\ref{constraint on joint
probabilities}) the CHSH inequality (\ref{CHSH}) using (\ref{delta}, \ref%
{p12&p15}) reduces itself to $\left\vert \Delta _{r}\right\vert \leq 1$\
where $\Delta _{r}=(p_{1}+p_{4}+3p_{5}-p_{8}+2p_{9}+2p_{14}-2)$. Now, if a
set of joint probabilities results in a NE in the quantum game and for this
set we have $\left\vert \Delta _{r}\right\vert \leq 1$\ then this NE has the
pseudoclassical aspect. However, if for this set we have $\left\vert \Delta
_{r}\right\vert >1$\ then it has a \textit{bona fide} quantum aspect. Note
that in the quantum MP game the strategy pairs $(1,0)\ $and $(1,1)$\ emerge
as NE for the set (\ref{first set}). For these NE we obtain $\Delta _{r}=2%
\sqrt{2}$. Similarly, the strategy pairs $(0,1)\ $and $(1,1)$\ emerge as NE
for the set (\ref{first set}) and for these NE we obtain $\Delta _{r}=-2%
\sqrt{2}$. These four\ NE, therefore, have a \textit{bona fide} quantum
aspect.

\begin{acknowledgement}
One of us (AI) is supported at the University of Adelaide by the Australian
Research Council under the Discovery Projects scheme (Grant No. DP0771453).
\end{acknowledgement}


\begin{thebibliography}{99}
\bibitem{FootNote1} Here by quantum mechanical probabilities we mean the
probabilities that are obtained from squaring the probability amplitudes. In
this paper we do not consider the negative probabilities that are sometimes
introduced to give another perspective on quantum phenomena.
\bibitem{FootNote2} As the two halves, which the players receive in a run,
are quantum mechanical objects, it is assumed that the referee is familiar
with quantum mechanics.
\bibitem{FootNote3} That is, in each run the referee's measurement generates
one of the $4$ possible pairs $(+1,+1),$ $(+1,-1),$ $(-1,+1),$ and $(-1,-1)$%
, where the first entry in a bracket is the measurement outcome along
Alice's chosen direction (which is either $S_{1}$ or $S_{2}$) and,
similarly, the second entry corresponds to Bob's chosen direction (which is
either $S_{1}^{\prime }$ or $S_{2}^{\prime }$).
\bibitem{FootNote4} It is to be noticed that the constraints (\ref%
{constraint on joint probabilities}) depend on the game being played and
also on the particular outcome that results for factorizable joint
probabilities.
\end{thebibliography}
\end{document}